\documentclass[a4paper,11pt]{article}

\usepackage{jinstpub} 
\usepackage{graphicx}
\usepackage{subfigure}
\usepackage{verbatim}
\usepackage{url}
\usepackage{lineno}

\title{\boldmath Design of High-speed readout electronics for the DarkSHINE electromagnetic calorimeter}

\author[a,b]{Y.H. Guo}
\author[c,a,b]{, S. Li}
\author[c,a,b]{, K. Liu}
\author[a,b]{, Y. Liu}
\author[a,b]{, Y.Q. Tan}
\author[a,b]{, J.N. Tang}
\author[a,b,1]{, W.H. Wu\note{Corresponding author.}}
\author[a,b,c]{, H.J. Yang}
\author[c,a,b]{, Z.Y. Zhao}
\author[a,b]{, W. Zhi}
\author[a,b]{and Z.Z. Zhou}

\affiliation[a]{Institute of Nuclear and Particle Physics, School of Physics and Astronomy, Shanghai Jiao Tong University, \\800 Dongchuan Road, Shanghai, China}

\affiliation[b]{Key Laboratory for Particle Astrophysics and Cosmology (MoE), Shanghai Key Laboratory for Particle Physics and Cosmology, \\800 Dongchuan Road, Shanghai, China}

\affiliation[c]{Tsung-Dao Lee Institute, Shanghai Jiao Tong Univeristy, \\1 Lisuo Rd, Shanghai, China}

\emailAdd{wuweihao@sjtu.edu.cn}

\abstract{The DarkSHINE experiment aims to search for dark photons by measuring the energy loss of the electrons recoiled from fixed-target. Its electromagnetic calorimeter is primarily responsible for accurately reconstructing the energy of the recoil electrons and bremsstrahlung photons. The performance of the electromagnetic calorimeter is crucial, as its energy measurement precision directly determines the sensitivity to the search for dark photons. The DarkSHINE electromagnetic calorimeter uses LYSO crystals to form a fully absorptive electromagnetic calorimeter. It utilizes SiPMs to detect scintillation light in the crystals, and its readout electronics system deduces the deposited energy in the crystals by measuring the number of photoelectric signals generated by the SiPMs. The DarkSHINE electromagnetic calorimeter aims to operate at an event rate of 1$\sim$10 MHz, detecting energies ranging from 1 MeV to 1 GeV. These aspects pose challenges to the readout electronics. To meet the requirements of high energy measurement precision, a high event rate, and a large dynamic range, we have researched and designed a readout electronics system based on dual-channel high-speed ADCs and a customized DAQ. The front-end amplification part of this system uses low-noise trans-impedance amplifiers to achieve high-precision waveform amplification. It successfully achieves a dynamic range up to a thousandfold through a double-gain readout scheme. The digital part uses 1 GSPS high-speed ADCs to achieve non-dead-time, high-precision waveform digitization. The DAQ part uses JESD204B high-speed serial protocol to read out the signal from ADC, and transmit it to PC software for processing and storage. Test results show a signal-to-noise ratio greater than 66 dBFS and an ENOB greater than 10.6 bits. Energy spectra measurements have been conducted using LYSO crystals and SiPMs, and an energy resolution of 5.96\% at the 2.6 MeV gamma peak of Th-232 has been achieved.}


\begin{document}
\maketitle
\flushbottom

\section{Introduction} 
In the past few decades, various astrophysical observations indicate that dark matter (DM) exists in the universe. DM does not appear to interact with the electromagnetic field, which means it is difficult to detect. In order to satisfy dark matter’s density observed by astronomy, DM mass in “freeze-out” is allowed from MeV to 10s TeV~\cite{serpico2004mev}. In the last decades, several underground experiments involving Xenon~\cite{aprile2019light}, LZ~\cite{aalbers2023first}, and CDEX~\cite{meng2021dark}, have been launched to search for the weakly interacting massive particles (WIMP) which in the GeV-TeV mass range. These experiments give strong limits to the DM in this mass range. The light DM particles in the MeV-GeV mass range have not been fully explored. The light DM particles is possible to interact with other DM particles via a new dark force. Dark photons mediate this dark force and can kinetically mix with photons to interact with standard model (SM) particles. Thus the dark photons serve as an important portal between DM and SM particles. Many accelerator-based experiments have been conducted to search for dark photons, such as NA64 at CERN~\cite{banerjee2019dark}, BESIII at BEPCII~\cite{ablikim2017dark}, Belle-II at SuperKEKB~\cite{abe2010belle}, and the LDMX experiment in R\&D~\cite{aakesson2018light}.

DarkSHINE~\cite{chen2023prospective} is an electron-on-fixed-target experiment that make use of a high repetition rate single electron beam provided by the SHINE linac~\cite{zhao2017two}. The electrons are accelerated to 8 GeV with a repetition of 1$\sim$10 MHz. It is hypothetically anticipated to produce dark photons via electrons and nuclei interacting dark bremstrahlungs as predicted by BSM theory. The dark photons then decay to DM candidates, which escape the detection with missing energy and missing momentum signatures. The challenge of the detector is to suppress background processes of SM and distinguish the dark photon signals with extreme small production rate. 

In this paper, we will introduce the electromagnetic calorimeter (ECAL) of DarkSHINE and the requirement of readout electronics system in section~\ref{sec:ECAL}. Then we will introduce our hardware design of the high-speed readout electronics in section~\ref{sec:hardware}, and the data acquisition (DAQ) design of the readout electronics in section~\ref{sec:DAQ}. In section ~\ref{sec:results}, we give the test results of performance evaluation and commissioning. Finally, we give a conclusion for such readout electronics design in section~\ref{sec:conclusion}.

\section{DarkSHINE ECAL and the requirement of readout electronics system} 
\label{sec:ECAL}
\begin{figure}[htbp]
	\centering
	\includegraphics[width=.55\textwidth]{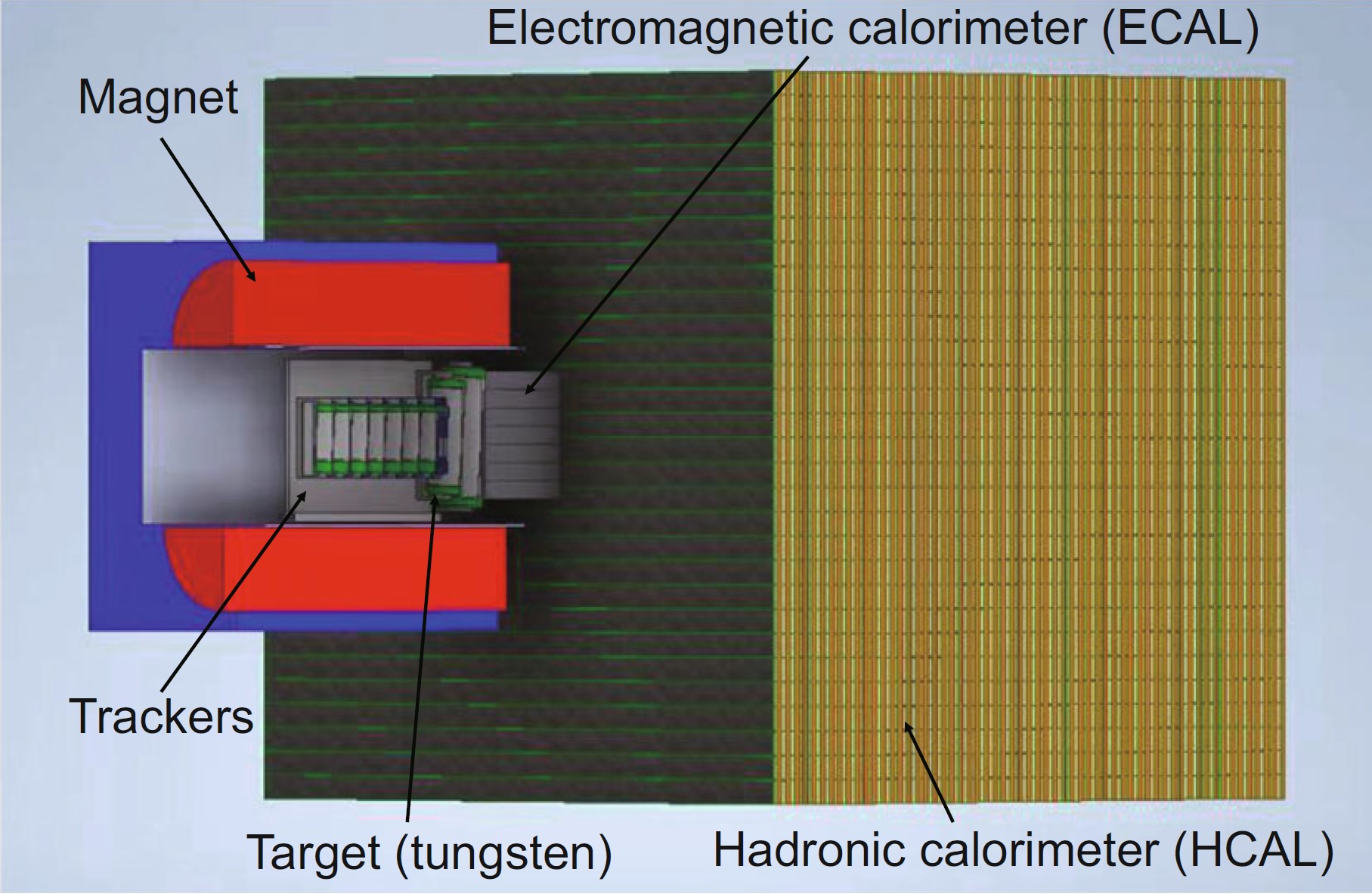}
	\caption{\label{fig:DS_detector}The detector sketch picture. The target is in the middle and the ECAL is placed after the recoil tracker.}
\end{figure}

The detector of DarkSHINE~\cite{li_2023_8373963}~\cite{dks} constitutes three sub-detector systems: silicon tracker, electromagnetic calorimeter (ECAL)~\cite{zzy}, and hadronic calorimeter (HCAL)~\cite{wang2024designhadroniccalorimeterdarkshine}, as shown in Figure~\ref{fig:DS_detector}. The ECAL is the key sub-detector of the system, which is placed after the silicon tracker to reconstruct the deposited energy of the incident particles, mainly responsible for accurately reconstructing the energy of the recoil electrons and bremsstrahlung photons. We implemented a crystal full absorption scheme for the ECAL, where the basic detection unit comprises LYSO crystals (light yield: 30,000 P/MeV, light attenuation time: 40 ns). The light yield is proportional to the energy deposited in the scintillator. By detecting the light yield in each collision event, the deposited energy could be reconstructed. 

The scintillation light will be converted into electric signals with low voltage, large dynamic range and fast rise time by silicon photomultiplier (SiPM), which has high photoelectron gain. Compared with other photoelectric conversion device, such as PMT or APD, SiPM has smaller size, better time resolution and lower cost, so we choose it as a photoelectric conversion device. The deposition energy range of a single crystal in the DarkSHINE electromagnetic calorimeter is 1 MeV to 1 GeV, and the number of photoelectric signals generated at SiPM is 100-100000, with the generated signals distributed in the range of about 200ns. This requires that the number of pixels in SiPM cannot be too small. After considering all SiPM models from Hamamatsu, we ultimately chose the S14160-3010PS model, with ~$1.8 \times 10^{5}$ gain, 89984 pixels. 


A readout electronics system of the ECAL is needed to process the photoelectron signals from SiPM and obtain the deposited energy in each event. 
Compared with other dark photon detection experiments, DarkSHINE experiment has very high event rate (1$\sim$10 MHz), thus the electronics system should process the SiPM signals within 100 ns. The rising edge of the SiPM signals is very fast, which requires a high-speed pre-amplifier to keep the waveform from distortion. Furthermore, the electronics system should have a dynamic range of approximately 1000 times to match the deposition energy range (1 MeV to 1 GeV) of the ECAL. DarkSHINE experiment requires an energy resolution of better than 1.5\% at 1 GeV, which poses a requirement for low-noise performance of the readout electronics system. Generally, the energy resolution of the ECAL can be expressed using the following formula:

\begin{equation}
  \frac{\sigma_{E}}{E(\rm{GeV})} = \frac{a}{\sqrt{E(\rm{GeV})}} \oplus \frac{b}{E(\rm{GeV})} \oplus c.
\end{equation}

\begin{figure}[htbp]
	\centering
	\includegraphics[width=.75\textwidth]{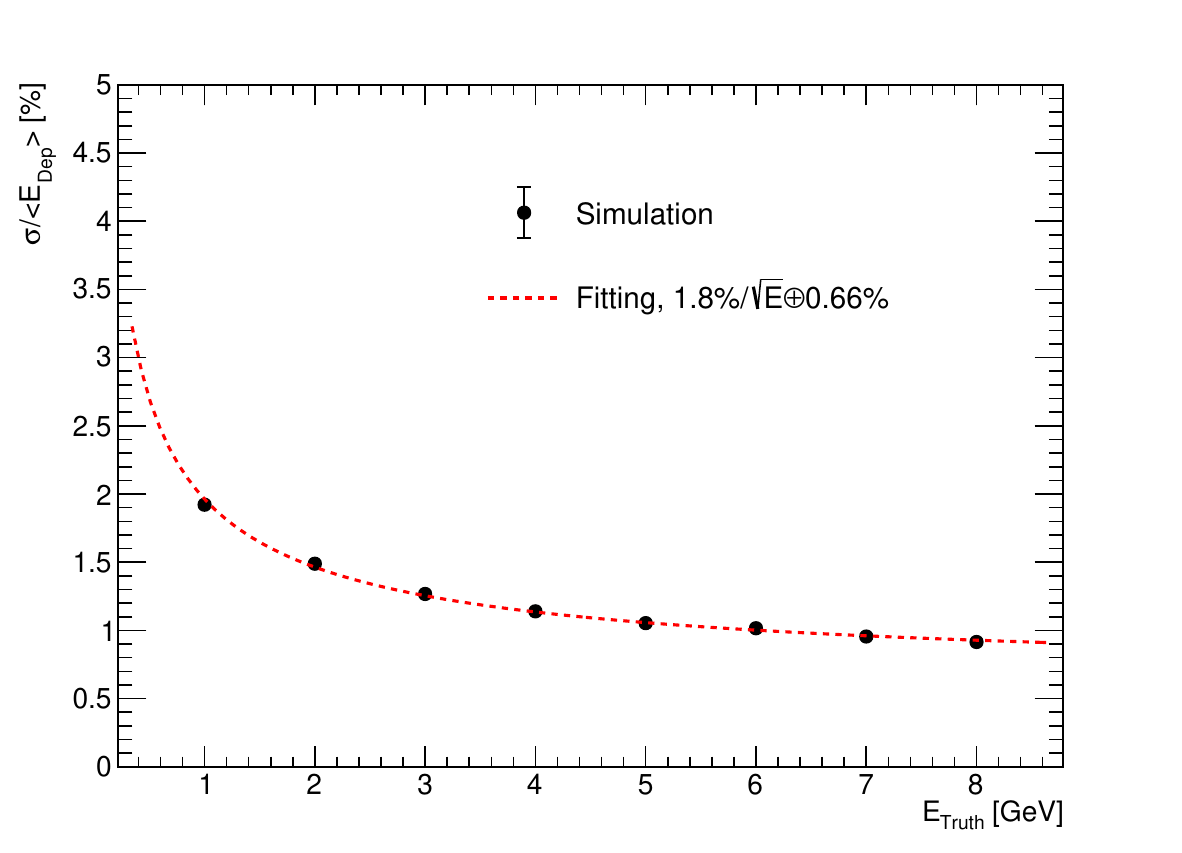}
	\caption{\label{fig:ECAL_ene_resolution}Energy resolution of the DarkSHINE ECAL given by Geant4 simulation.}
\end{figure}

In the formula, $a$ is the random term, mainly affected by statistical fluctuations, $b$ is the noise term, mainly affected by electronic noise, and $c$ is the constant term, mainly affected by the non-uniformity of the detector and the uncertainty of the scale. We obtain the energy resolution of the DarkSHINE ECAL through simulation with Geant4, as shown in Figure~\ref{fig:ECAL_ene_resolution}. In this experiment, $a = 1.2\%$ and $c = 0.57\%$, so $b$ should be less than $0.7\%$, equivalent to 7 MeV noise energy at 1 GeV energy mesurement.

\section{Hardware design of the high-speed readout electronics} 
\label{sec:hardware}

\begin{figure}[htbp]
	\centering
	\includegraphics[width=.75\textwidth]{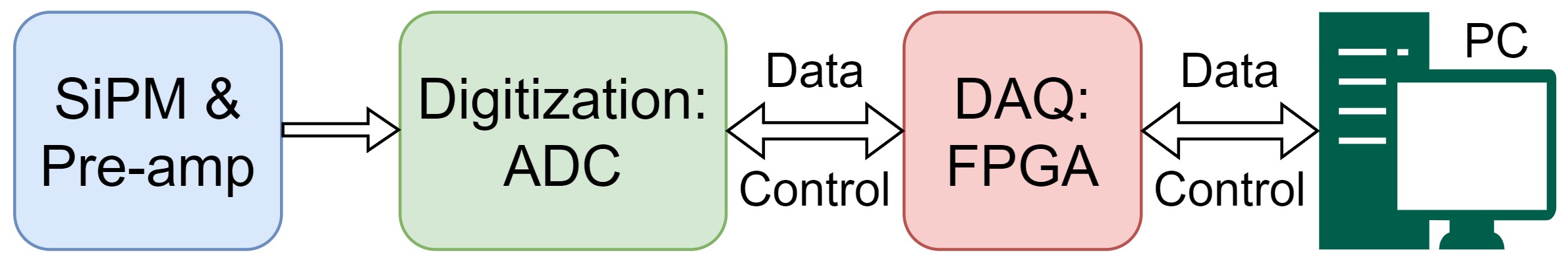}
	\caption{\label{fig:simple_block}The block diagram of the readout electronics system for DarkSHINE ECAL.}
\end{figure}

The readout electronics system is responsible for processing the photoelectron signals generated by SiPMs. As shown in Figure~\ref{fig:simple_block}, the signals are first preamplified and shaped. After being digitized by a analog-to-digital converter (ADC), the data will be transferred to the DAQ system for processing and then transferred to computer via Ethernet. 

\subsection{Design of the pre-amplifier board}
The electrical signal converted by SiPM is a small amplitude current pulse signal, which needs to be converted into a voltage pulse signal and amplified. The output signal waveform width of SiPM is greater than 200ns, which is larger than the signal repetition interval of 100ns. Therefore, it is necessary to perform shaping filtering on it to avoid signal accumulation.

Generally, charge sensitive amplifiers (CSA)~\cite{manghisoni2015dynamic} or Trans-impedance amplifiers (TIA)~\cite{grybos2006integrated} are used as front-end amplifiers for I/V conversion. CSA may lose signal waveforms and have longer output trailing edges, while TIA can avoid these issues, so we ultimately adopt TIA as front-end amplifier. The rising edge width of SiPM signal is about 5ns, corresponding to a bandwidth of 70MHz. If the bandwidth of TIA is not wide enough, it will cause signal distortion. The bandwidth of the TIA is calculated according to equation~(\ref{eq:TIA_BW})
\begin{equation}
	F_{{\rm -3dB}}=\sqrt{\frac{\rm GBP}{2\pi \cdot R_{\rm F} \cdot C_{\rm S}}} \label{eq:TIA_BW}
\end{equation}
GBP represents the gain bandwidth product of the operational amplifier. $R_{\rm F}$ reprensents the trans-impedance gain resistor, with a value of 1 k$\Omega$.
$C_{\rm S}$ represents the total parasitic capacitance at the inverting input, mainly the parasitic capacitance of the SiPM which is positively correlated with the photosensitive area. To obtain large bandwidth and improve low-noise performance, the amplifier LMH6629 with 4.0 GHz GBP is applied to construct the TIA. We get a bandwidth of $\sim$150 MHz of the TIA at the gain of 1k.

\begin{figure}[htbp]
    \subfigure[Schematic of the pre-amplifier board.]{
            \centering
  	    \includegraphics[width=1\textwidth]{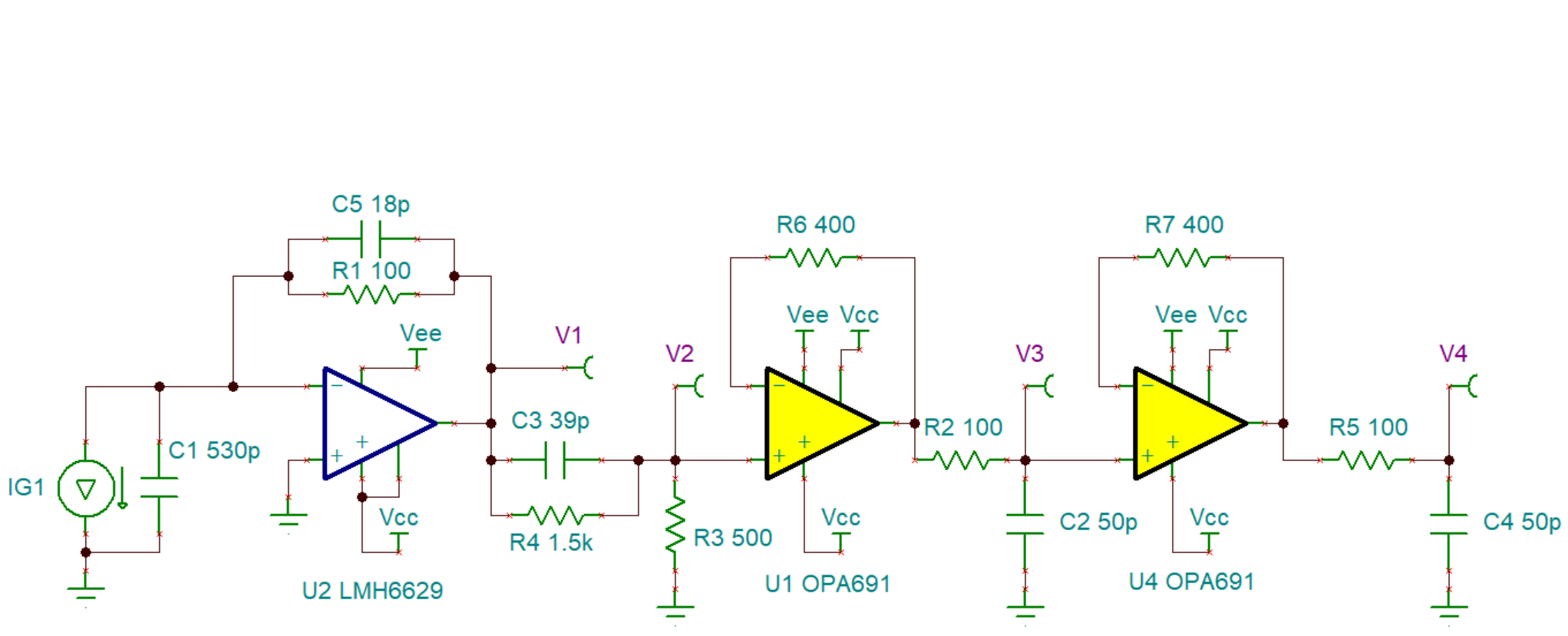}
   	    \label{fig:preamp_sch}
    }
    \newline
    \subfigure[A photograph of the pre-amplifier board.]{
            \centering
  	    \includegraphics[width=.42\textwidth]{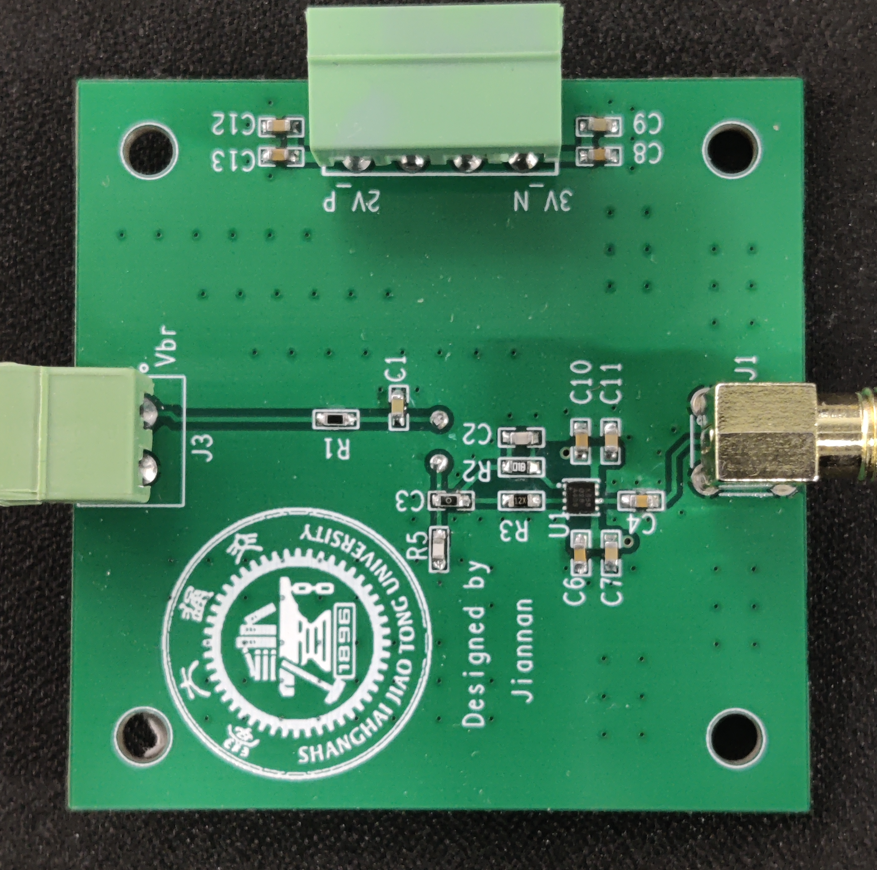}
   	    \label{fig:preamp_photo}
    }
    \hfil
    \subfigure[A photograph of the dual-gain amplifier.]{
            \centering
  	    \includegraphics[width=.48\textwidth]{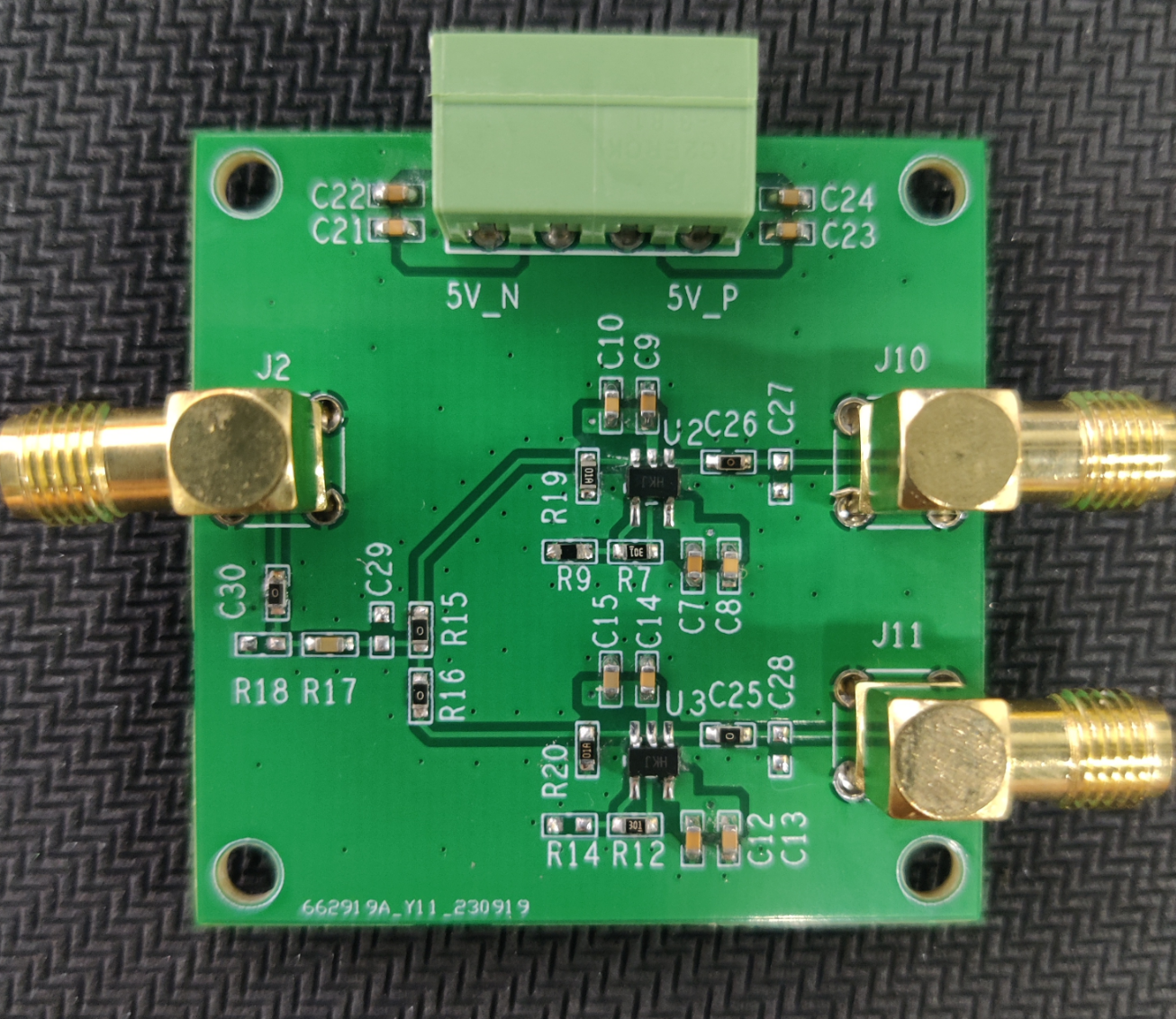}
   	    \label{fig:2gain_photo}
    }
    \caption{\label{fig:preamp}Schematic and photograph of the pre-amplifier board.}
\end{figure}

To narrow the output signal width and meet the requirement of a 100ns event interval, it is also necessary to filter the signal. We adopte a “TIA-CR-$\rm RC^{2}$” structure, shown in Figure~\ref{fig:preamp_sch}. 
Each stage of the filtering circuit is isolated by a voltage follower composed of an operational amplifier. The CR circuit can filter out low-frequency components and narrow the trailing edge waveform, and the two-stage RC circuit can filter out high-frequency noise, making the front edge of the output waveform smoother, closer to Gaussian, and has a better signal-to-noise ratio. The photograph of the pre-amplifier board is shown in Figure~\ref{fig:preamp_photo}.

\begin{figure}[htbp]
	\centering
	\includegraphics[width=.75\textwidth]{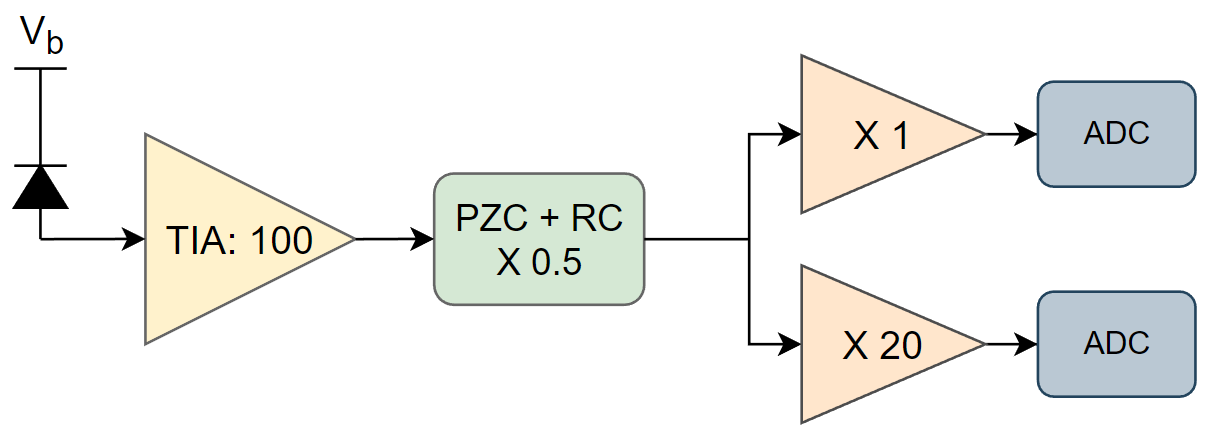}
	\caption{\label{fig:gain}Dual channel high and low gain scheme.}
\end{figure}

The energy range of DarkSHINE is 1MeV-1GeV, and the required dynamic range is about 10 dB. In order to achieve larger dynamic range and better energy resolution, we adopt a dual channel high and low gain scheme. The dual-gain amplifier that is placed in front of the ADC is set to high gain (20x) when the input energy is lower than a certain threshold level (about 40MeV) to improve the SNR. To avoid exceeding the ADC's input range, the gain is set down to low gain (1x) when the input energy is higher than the threshold. The scheme of the dual-gain amplifier is shown in Figure~\ref{fig:gain}, and a photograph of it is shown in Figure~\ref{fig:2gain_photo}.

\subsection{Design of the ADC board}
After the SiPM analog signal is amplified and shaped, high-precision digitization is also required for signal processing in digital circuits. In order to meet the requirements of high sampling rate, high precision and large dynamic range, A dual-channel 14-bit, 1 GSPS ADC (AD9680) is employed for waveform digitization.

\begin{figure}[htbp]
    \subfigure[Block diagram of ADC board.]{
            \centering
  	    \includegraphics[width=.4\textwidth]{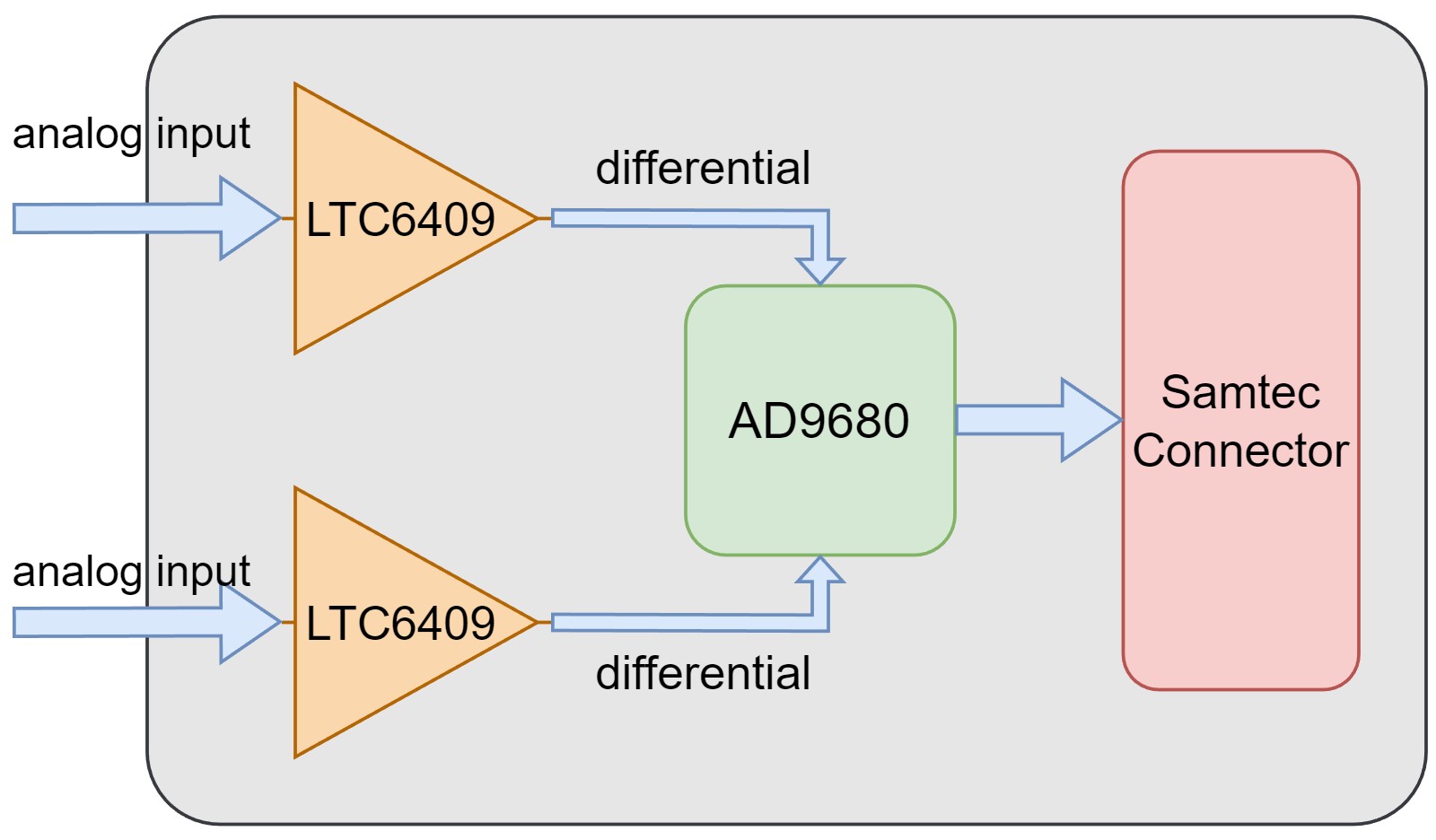}
   	    \label{fig:adc_block}
    }
    \hfil
    \subfigure[A photograph of the ADC board]{
            \centering
  	    \includegraphics[width=.4\textwidth]{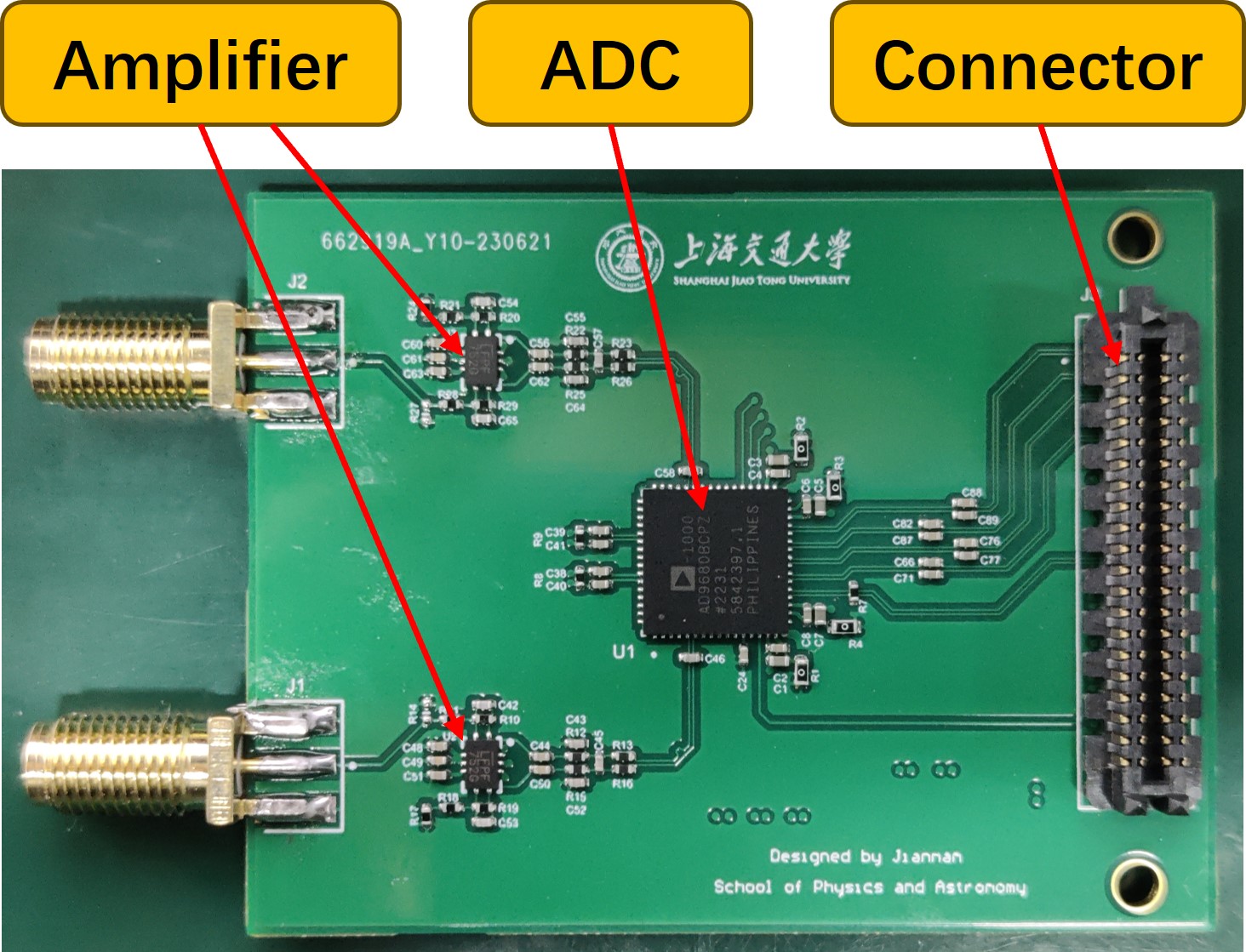}
   	    \label{fig:adc_photo}
    }
    \caption{\label{fig:adc}Block diagram and photograph of the ADC board.}
\end{figure}

We designed an ADC board based on AD9680, as shown in Figure~\ref{fig:adc}. The ADC board mainly contains 2 single end to differential drives (LTC6409), 1 dual channel ADC chip (AD9680) and a 20 * 4 pins board-to-board connector (Samtec). The required power and clock are provided by FPGA board, which will be discussed in the following section.

\subsection{Design of the FPGA board}
\begin{figure}[htbp]
	\centering
	\includegraphics[width=.7\textwidth]{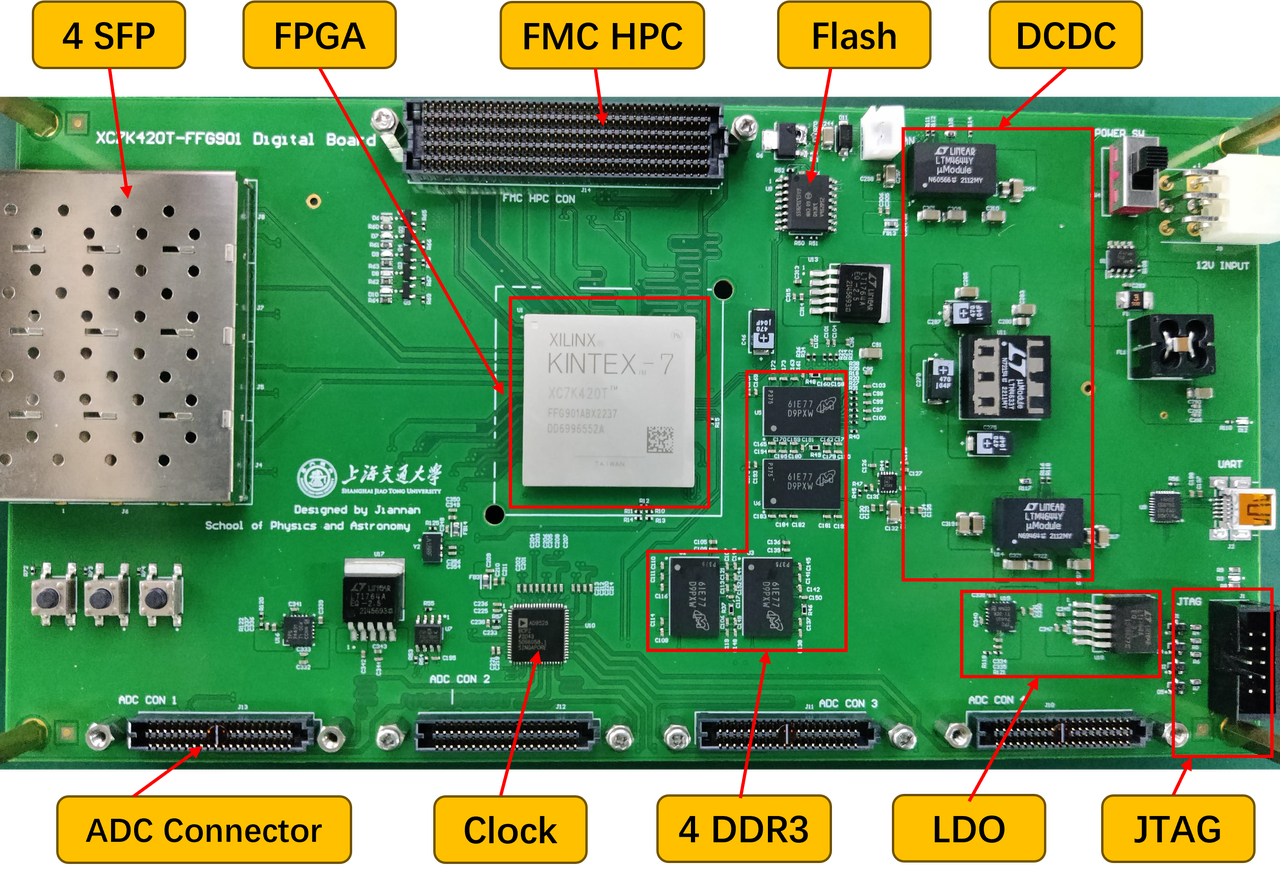}
	\caption{\label{fig:FPGA}A photograph of the FPGA board.}
\end{figure}

The signal digitized by ADC will be transmitted to FPGA for digital processing. An FPGA board is connected with four ADC sub-board, and the output of one ADC is divided into four lanes based on JESD204B high-speed serial protocol~\cite{saheb2014scalable}, with normal working data rate of 10 Gbps for each lane. 
We choose XC7K420T-FFG901 from Xilinx Kindex-7 series as the FPGA. XC7K420T-FFG901 has 28 GTX channels for JESD204B data transmission, which supports serial data transmission at a maximum speed of 12.5Gbps. The photograph of the board designed around XC7K420T-FFG901 is shown in figure ~\ref{fig:FPGA}. The FPGA board mainly contains 4 ADC connectors for ADC connection, 4 SFP connectors for 10G Ethernet transmission, 1 256Mb Flash for configuration file storage, 1 JTAG connector for online configuration, 4 512Mb DDR3 for high-capacity data storage, 1 FMC HPC for expanding other applications. There is also a clock system and a power system on the board. SIT9121 provides 200 MHz system clock for FPGA, and AD9528 provides clock for ADC sub-boards and FPGA used in JESD204B protocol. 
The design of DCDC + LDO provide the ADC high-quality power.

\section{DAQ design of the high-speed readout electronics}
\label{sec:DAQ}

\subsection{Design of the Firmware}
\begin{figure}[htbp]
	\centering
	\includegraphics[width=.9\textwidth]{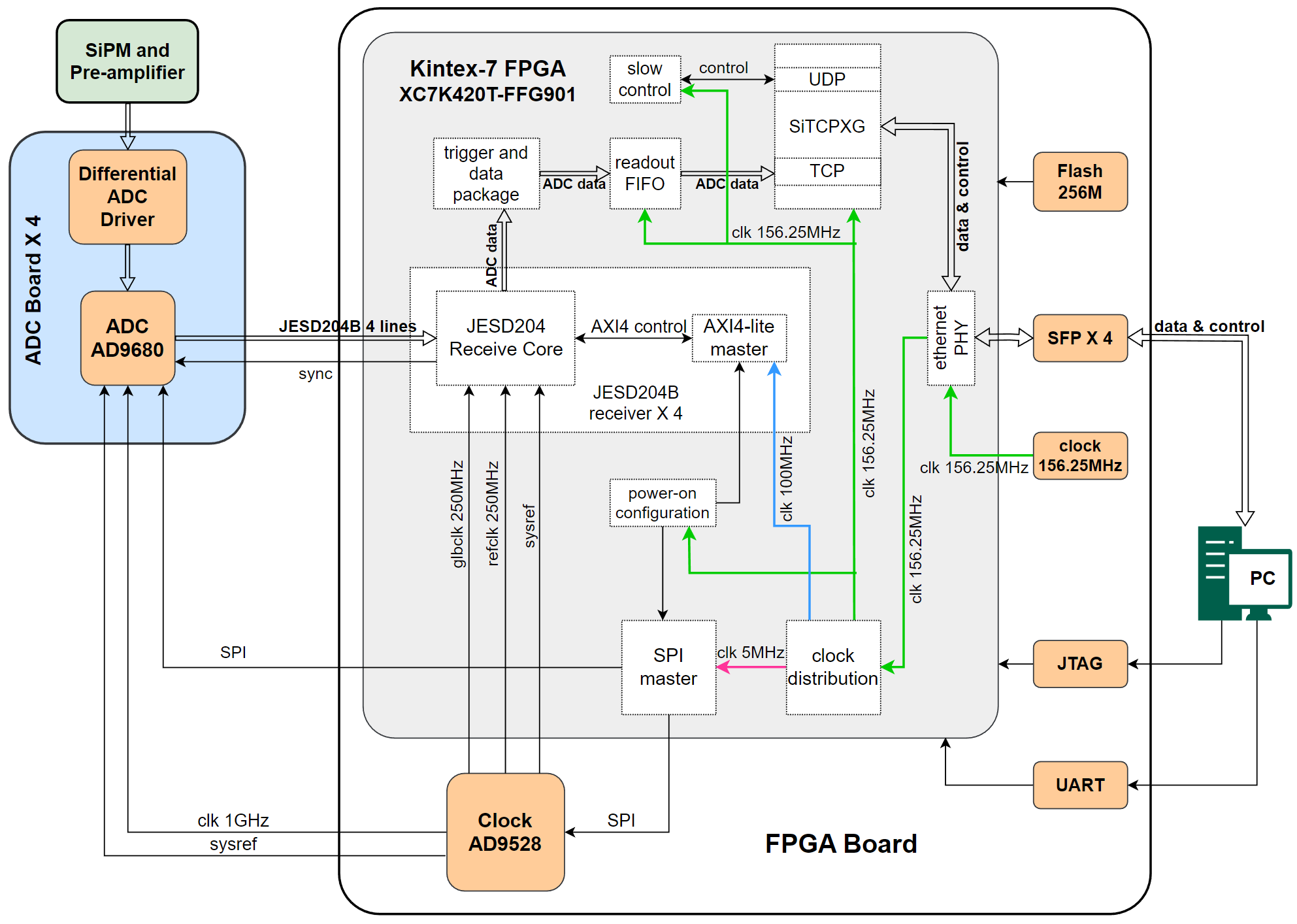}
	\caption{\label{fig:FPGA_block}The block diagram of the firm ware design for the readout electronics.}
\end{figure}

The firmware running in the FPGA is developed by the Xilinx’s Vivado using Verilog HDL programs, guidding the work of the whole electronic system. The block diagram of the firmware is shown in Figure~\ref{fig:FPGA_block}. The firmware mainly includes JESD204B data transmission wrapper, ADC data triggering and processing wrapper and ethernet communication wrapper.

JESD204B data transmission wrapper: This module is responsible for receiving the high speed serial data from ADC, and translating it into parallel output. The input data of each ADC channel has a sampling rate of 1 Gsps. Each sample is digitized to 14 bits, and the output of each channel is divided into two lanes. For one sample, Bit 13 to Bit 6 will be distributed to the first lane, and Bit 5 to Bit 0 will be supplemented with 0 to 8 digits, and distributed to the second lane. Afterwards, data from each lane will be 8b / 10b encoded, serialized, and then input into FPGA via the SerDes interface, with a data rate of 10Gbps. We use JESD204 IP core in the Vivado to handle the data. One JESD204 IP receives two ADC channels' data, and decodes the 1GHz, 4 * 10 bits serial input data into 250MHz, 128 bits parallel output data. The synchronization of multiple ADCs is achieved by performing AND operations on the SYNC signal output from the FPGA, and then connecting it to each ADC separately.

ADC data triggering and processing wrapper: The output of JESD204 IP for each ADC channel will enter a ring buffer for temporary storage. When the buffer is full, the earliest data will be replaced by new the data, just like a ring. A trigger system is designed for the ring buffer, that when a sample value exceeds the set threshold, a certain length of samples before and after it, along with a header containing channel information, trigger count information, and trigger time information, will be sequentially output to the next level. All of these samples can be regarded as a SiPM threshold-crossing waveform.The offset, length, and threshold of the trigger system can all be configured by PC software.

Ethernet communication wrapper: SiTCP is the technology to connect a physical experiment front-end to PC via Ethernet~\cite{uchida2008hardware}. There are two types of user interface to connect the SiTCP and the user circuit, TCP interface and UDP interface. We use TCP interface for data transmission, and UDP interface for slow control.

\subsection{Design of the PC Software}
\begin{figure}[htbp]
	\centering
	\includegraphics[width=.8\textwidth]{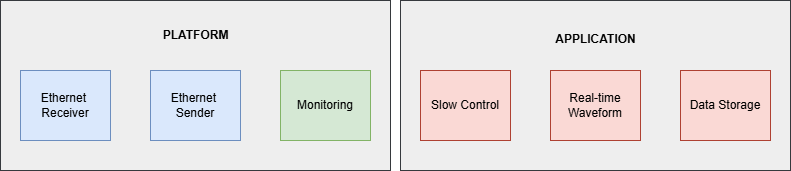}
	\caption{\label{fig:PC_software_block}Block of PC Software.}
\end{figure}

The PC software is responsible for sending commands and receiving data, which is developed based on PyQt5. As shown in Figure~\ref{fig:PC_software_block}, PC software is structured in the platform layer and application layer.

The platform layer contains functions for network and monitoring. Ethernet receiver function uses TCP protocol to receive the data transmitted from FPGA, and ethernet sender function uses UDP protocol to send the slow control configuration to configure the register in FPGA. The monitoring function helps users to monitor the link state, sample rate, receiving count and ADCs' sync state.

\begin{figure}[htbp]
	\centering
	\includegraphics[width=.9\textwidth]{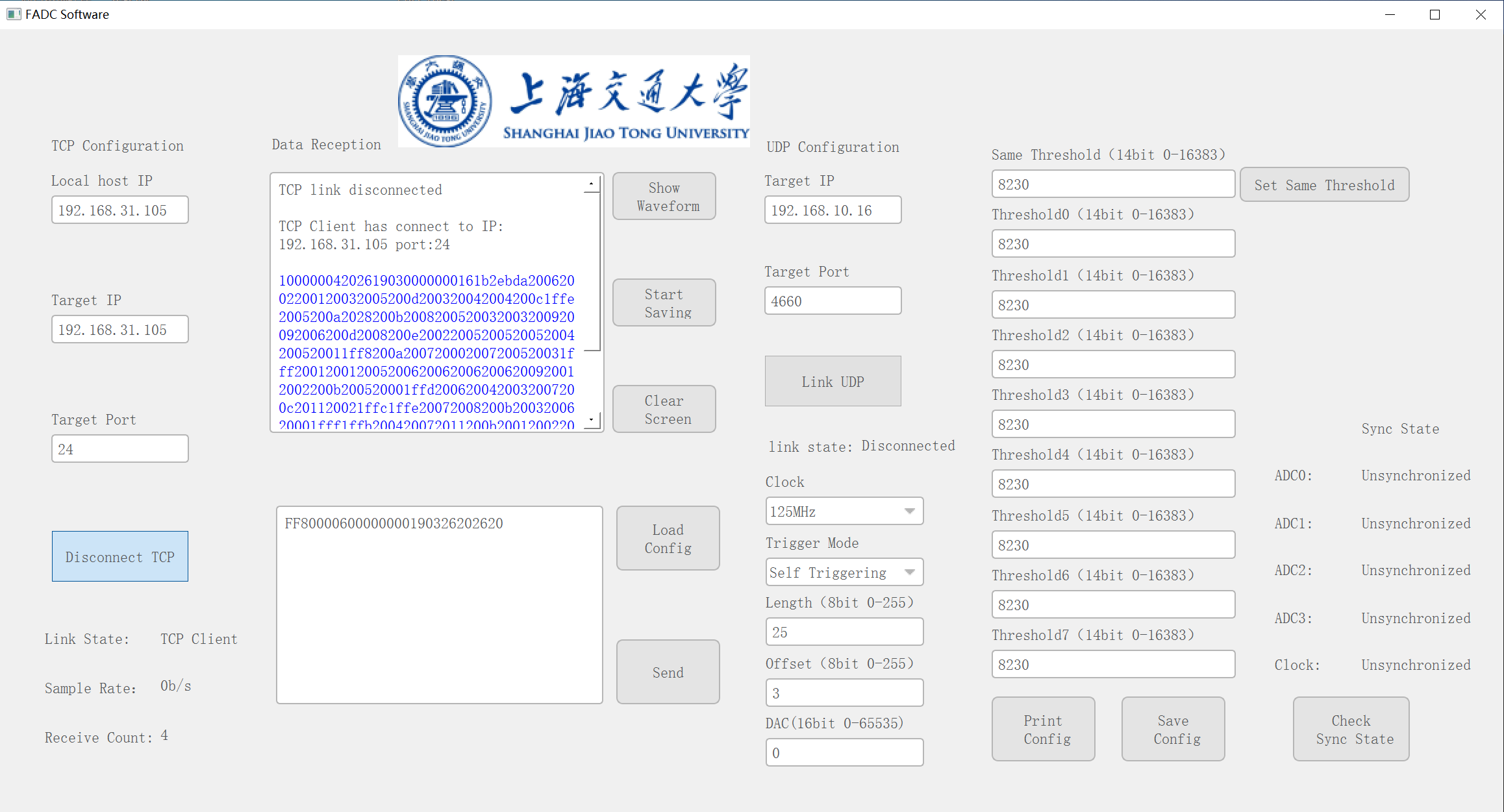}
	\caption{\label{fig:PC_software}A screenshot of PC Software.}
\end{figure}

The application layer provides users high-level functions. The slow control part provides users a GUI interface to configure the slow control registers in FPGA. Users can also see real-time waveform images of FPGA transmitted data and store these data locally through the graphical interface provided by this layer. A screenshot of the PC software is shown in Figure~\ref{fig:PC_software}.

\section{Results}
\label{sec:results}

\subsection{Performance evaluation test}
\begin{figure}[htbp]
	\centering
	\includegraphics[width=.55\textwidth]{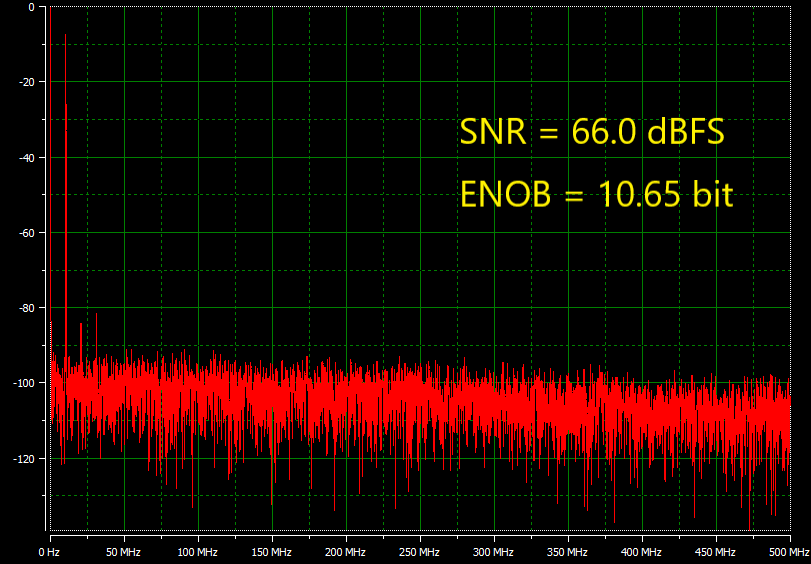}
	\caption{\label{fig:FFT_10P3M}The frequency spectrum of the ADC output with the input of 10.3 MHz sine wave.}
\end{figure}

We have conducted the performance evaluation tests using a signal source (Tektronix AFG31252). The ADC input is a standard 10.3 MHz sine wave with an amplitude close to the full scale of the 1.7V peak-to-peak ADC input. The fast Fourier transform (FFT) is applied to convert the ADC output from the time domain to the frequency domain. We use $2^{14} = 16384$ ADC samples as the input to the FFT algorithm, obtaining the frequency spectrum and the performance indicators. As shown in Figure~\ref{fig:FFT_10P3M}, the readout system has an SNR of approximately 66 dBFS for both channels, with an ENOB of about 10.6 bits.

\subsection{Commissioning test with Th232 radiation source}
\begin{figure}[htbp]
	\centering
	\subfigure[The block diagram of the test platform]{
  	    \includegraphics[width=.65\textwidth]{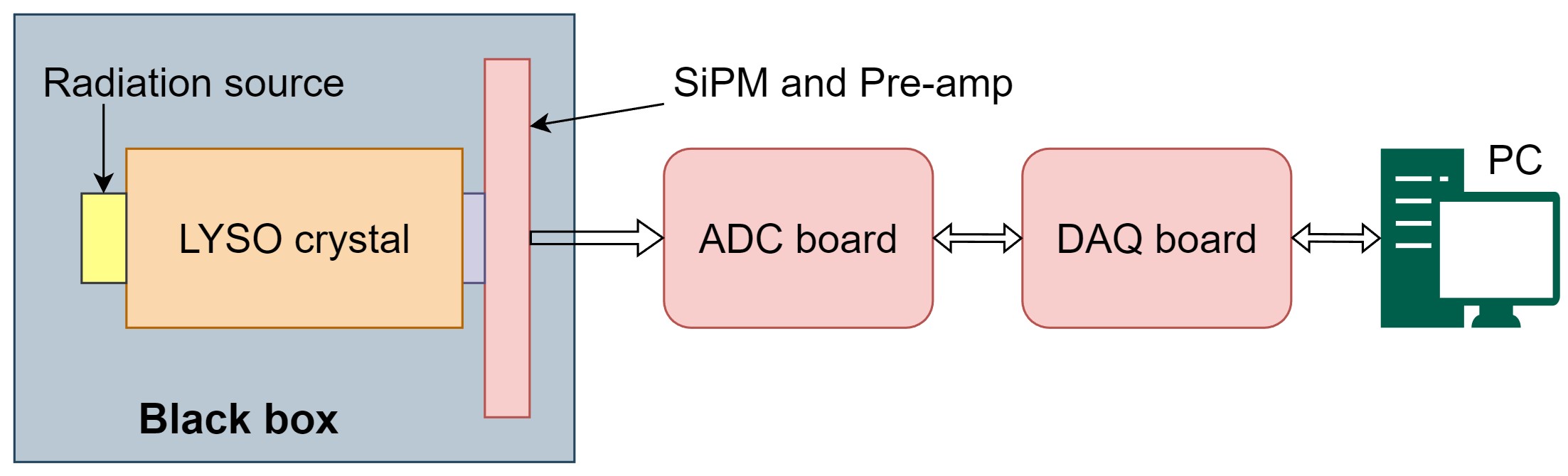}
   	    \label{fig:test_block}
    }
    \subfigure[A photograph of the test platform.]{
  	    \includegraphics[width=.6\textwidth]{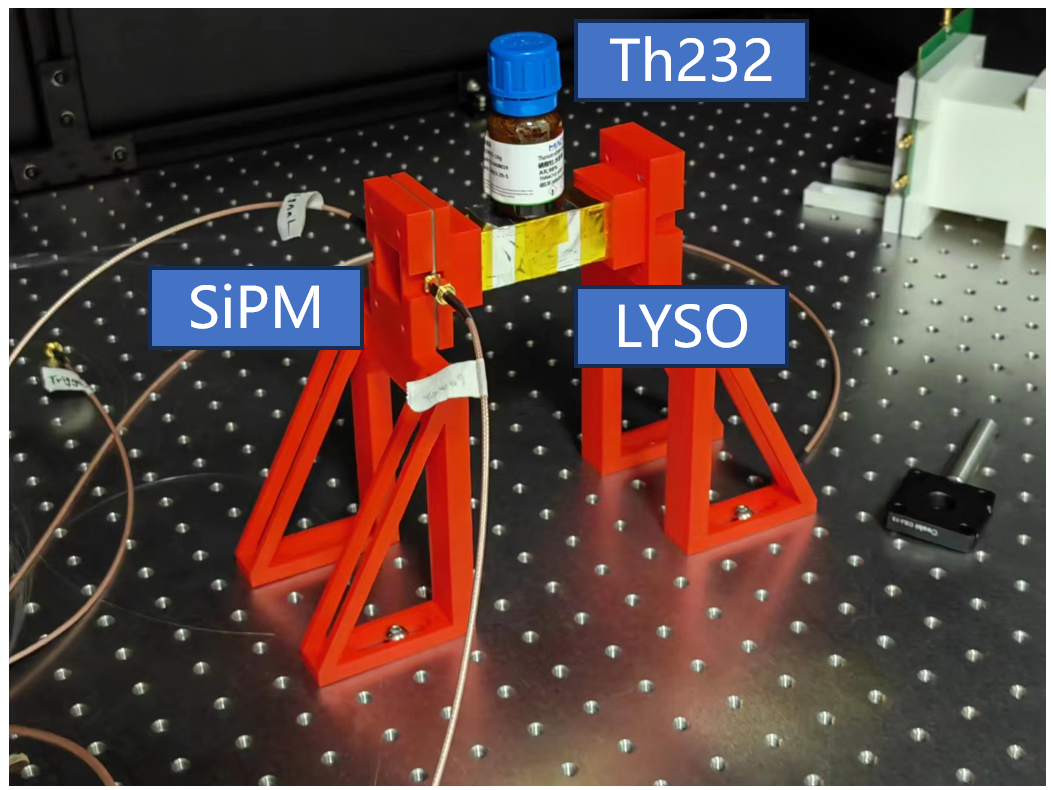}
   	    \label{fig:platform_photo}
    }
    \caption{\label{fig:platform}Block diagram and photograph of the test platform  using Th-232 radiation source.}
\end{figure}

The commissioning test of the readout electronics with LYSO crystals has also been conducted. The block diagram and a photograph of the test platform is shown in Figure~\ref{fig:platform}. The detector section consists of a Th-232 radiation source, which emits $\gamma$ rays with an energy of 2.6 MeV, an LYSO crystal, and S14160-3010PS SiPM. 

LYSO crystal absorbs $\gamma$ rays and emits light. The emitted light is converted into electrical signals by SiPM, amplified by the front-end, and digitized by ADC. The digital signal is processed by FPGA and finally output to the PC software.


\begin{figure}[htbp]
	\centering
    \subfigure[The SPE spectrum of the S14160-3010PS SiPM.]{
  	    \includegraphics[width=.48\textwidth]{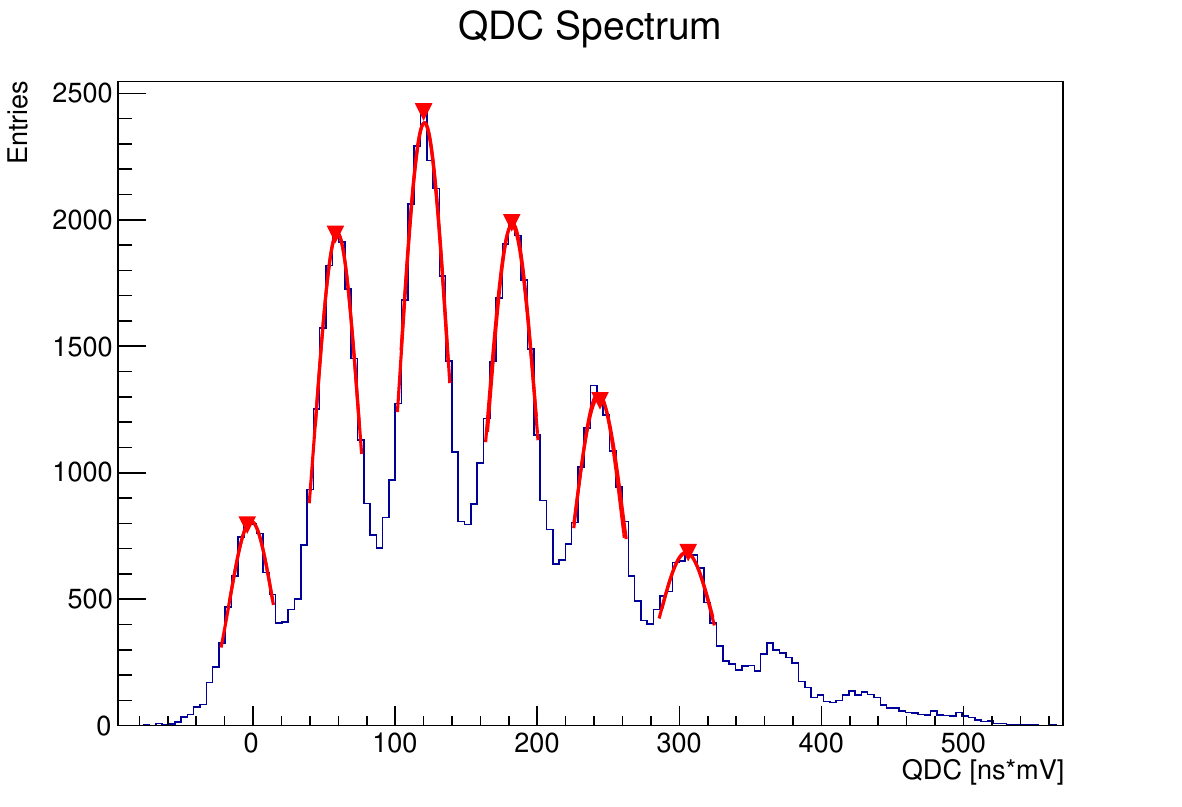}
   	    \label{fig:SPE spectrum}
    }
    \hfil
    \subfigure[The linear fit of the SPE spectrum for the S14160-3010PS SiPM.]{
  	    \includegraphics[width=.48\textwidth]{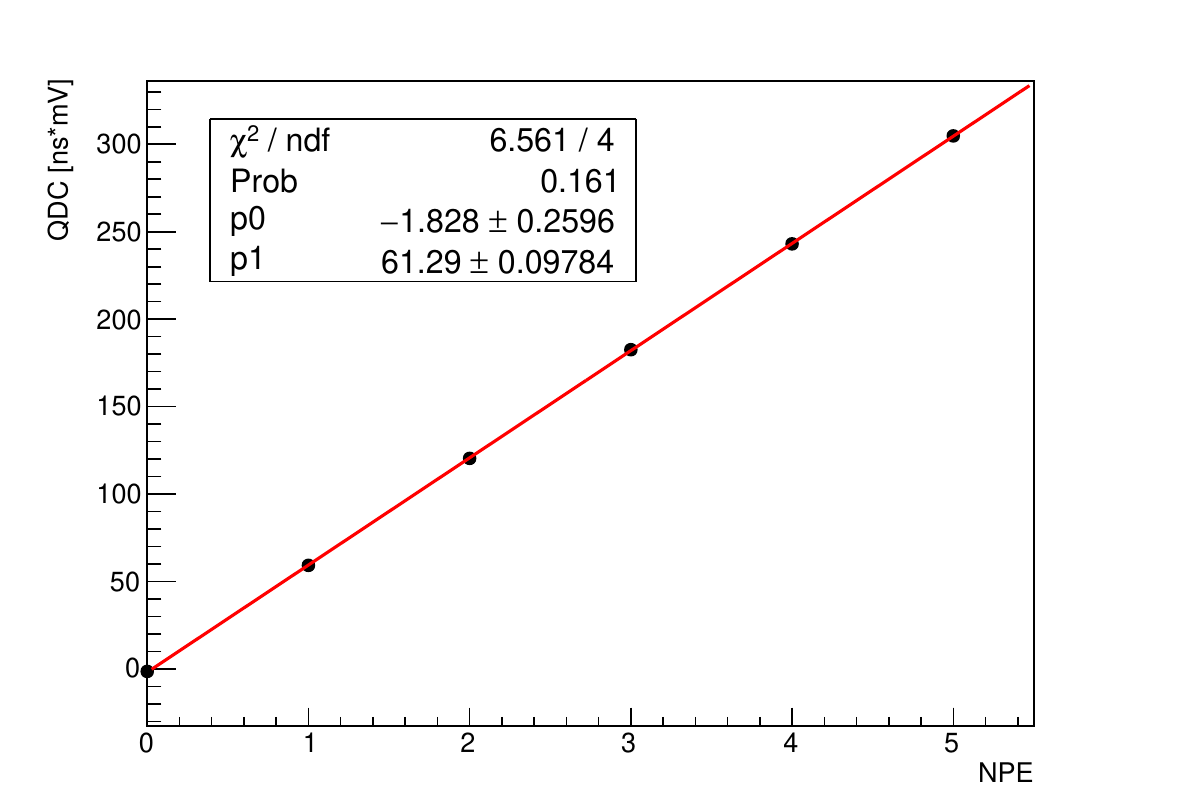}
   	    \label{fig:SPE fit}
    }
    \caption{\label{fig:spe}The calibration result of S14160-3010PS SiPM.}
\end{figure}

Before starting the test, it is necessary to use LED calibrating the SiPM gain, and fit the single photoelectron spectrum. The calibration result is shown in Figure~\ref{fig:spe}.

\begin{figure}[htbp]
	\centering
	\includegraphics[width=.7\textwidth]{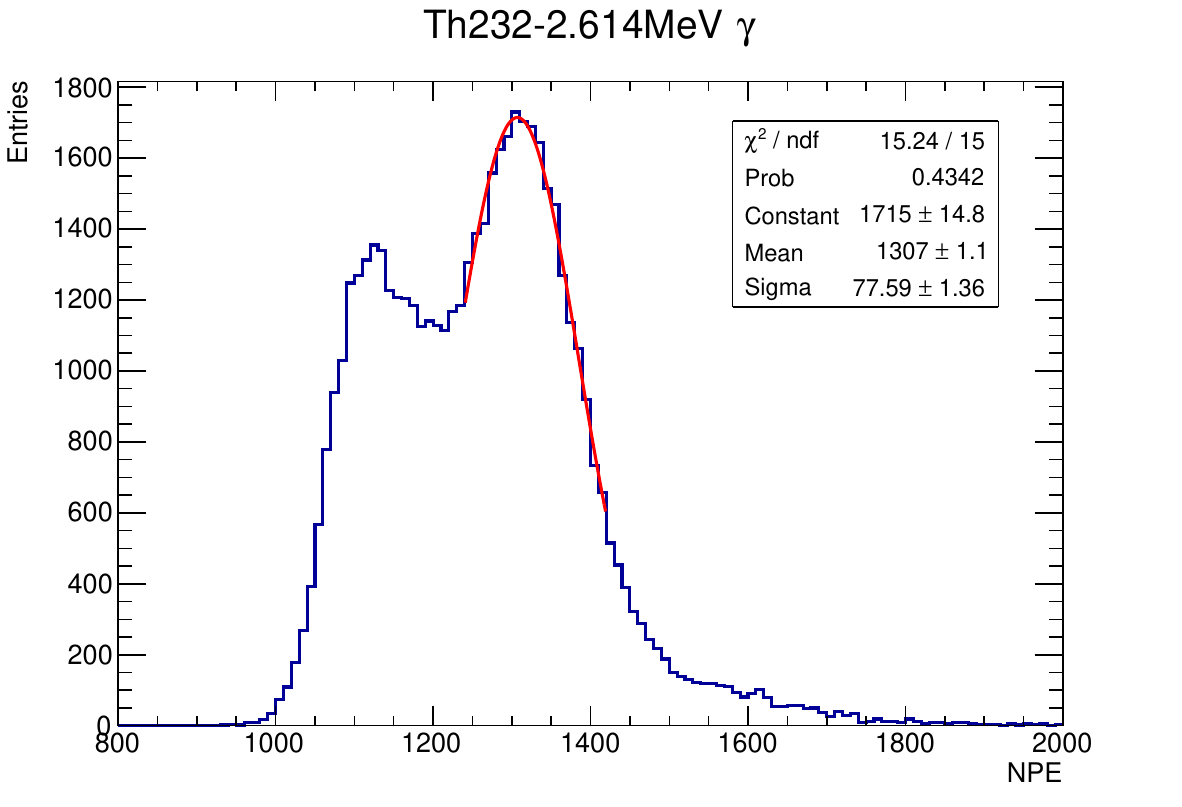}
	\caption{\label{fig:Th-232}The spectrum of Th-232 2.6 MeV $\rm \gamma$ ray using S14160-3010PS at 1 k$\Omega$ transimpedance gain.}
\end{figure}

The energy spectrum of 2.6 MeV $gamma$ rays from Th-232 obtained by integrating the output waveform of SiPM is shown in Figure~\ref{fig:Th-232}. Through Gaussian fitting, the detector used in this test has an energy resolution of 78.8/1307=6\% when detecting 2.6 MeV gamma rays.

\section{Conclusion}
\label{sec:conclusion}
DarkSHINE is a future fixed-target experiment with a event rate of 1$\sim$10MHz designed to search for dark photons. The DarkSHINE ECAL reconstructs the energy of the incident particles with LYSO crystals and SiPMs. To process the fast SiPM signals, the high-speed readout electronics with large analog bandwidth, high sample rate and large dynamic range are designed and tested.
The pre-amplifier board has a structure of “TIA-CR-$\rm RC^{2}$” together with dual channel high and low gain, amplifying the signal and improving the SNR.
The sub-board with dual channel, 1 GSPS sample rate ADC as the core is designed to digitize the SiPM signals. The SNR of the ADC is 66 dBFS, and the ENOB is 10.6 bits, fulfilling the design requirements. 
A digital FPGA board and a PC software have also been designed for digital processing. We obtain the energy spectrum of 2.6 MeV $gamma$ rays from Th-232 with our readout electronics and each part of the electronics works stable.
These tests results validate the readout electronics meet the requirement of the DarkSHINE experimental conditions in terms of fast signal readout and data acquisitions.

\acknowledgments
This work was supported by the Ministry of Science and Technology of China (No.: 2023YFA1606203), Shanghai Pilot Program for Basic Research — Shanghai Jiao Tong University (No.: 21TQ1400218 and 21TQ1400209), Yangyang Development Fund, National Key R\&D Program of China (Grant No.: 2023YFA1606904 and 2023YFA1606900), and National Natural Science Foundation of China (Grant No.: 12150006).








\bibliographystyle{JHEP}
\bibliography{myref}
\end{document}